\documentclass[preprint,nofootinbib,prd,aps]{revtex4}
\usepackage{graphicx,amstext,amssymb,amsmath,color,ulem,hyperref}
\begin{document}

\title{Gradient corrections to the local-equilibrium energy-momentum tensor in the 
Zubarev approach}

\author{Sergii V. Akkelin}

\affiliation{Bogolyubov Institute for Theoretical Physics,
Metrolohichna  14b, 03143 Kyiv,  Ukraine }
\author{Dirk H.~Rischke}
\affiliation{Institut f\"ur Theoretische Physik,
	Johann Wolfgang Goethe--Universit\"at,
	Max-von-Laue-Str.\ 1, D--60438 Frankfurt am Main, Germany} 
\affiliation{Helmholtz Research Academy Hesse for FAIR, Campus Riedberg,\\
	Max-von-Laue-Str.~12, D--60438 Frankfurt am Main, Germany}

\begin{abstract}

We compute the expectation value of the energy-momentum tensor of a real scalar field in an approximation which accounts for spacetime gradients of the hydrodynamical variables in local thermodynamical equilibrium. 
We show that the energy-momentum tensor receives corrections with respect to the standard local-equilibrium result.
Notably, the relation between the energy density and pressure, i.e., the equation of state, is modified with respect to the one in global equilibrium.
The obtained corrections might be relevant for systems created in relativistic hadron and heavy-ion collisions.

\end{abstract}


 \maketitle

\section{Introduction}

Relativistic hydrodynamics (see Ref.~\cite{Rischke} for a recent review of theories of relativistic dissipative hydrodynamics)  is successfully applied at very different scales: from hadron and heavy-ion collisions \cite{Florkowski,Romatschke,Shen} to astrophysics \cite{Mukhanov,Rez}. 
The applicability of hydrodynamics indicates the validity of some reduced description for a given system. 
Here, ``reduced description'' means that the state of a system can be characterized by the knowledge of the expectation values of some observables only.
It has been well known for a long time (see  Ref.~\cite{Jaynes}) that the statistical operator $\rho$ of a state which is ``least biased'' as far as unmonitored degrees of freedom are concerned maximizes the von Neumann entropy, $\mathcal{S}=-\rm{Tr} [\rho \ln{\rho}]$, subject to  the constraints  
\begin{eqnarray}
\langle A_{n} \rangle = \mathrm{Tr} [A_{n} \rho]\;, \label{1} \\
\mathrm{Tr} [ \rho ]=1\;, \label{2}
\end{eqnarray}
where  $\langle A_{n} \rangle $ are the expectation values of some (relevant)  observables (operators) $A_n$.
The statistical operator is then given by that of a generalized Gibbs state 
\begin{eqnarray}
\rho=\frac{1}{Z}\exp \left ( - \sum_{n} a_{n} A_{n} \right )\;,
\label{3} 
\end{eqnarray}
where $a_{n}$ are the corresponding Lagrange multipliers and 
\begin{eqnarray}
Z=\mathrm{Tr}\left[\exp \left ( - \sum_{n} a_{n}  A_{n} \right )\right]
\label{4}
\end{eqnarray}
is the partition function which, as a normalizing factor in Eq.~\eqref{3}, ensures that $\mathrm{Tr}[\rho]=1$. 

The crucial point for the applicability of the reduced description is the choice of the set of relevant observables. 
For a quantum field-theoretical system in a heat bath of temperature $T$, without a conserved charge and in global equilibrium, the sole relevant observable is the Hamilton operator $H$, and the statistical operator reads in the rest frame of the system
\begin{eqnarray}
\rho_{\text{geq, RF} } = Z^{-1}_{\text{geq, RF}} \exp\left(-\beta H \right)\;, \quad
Z_{\text{geq, RF}} = \mathrm{Tr} \left[ \exp\left(-\beta H \right) \right] \;,\label{rho_geqRF}
\end{eqnarray}
where $\beta \equiv 1/T$.\footnote{We use natural units $\hbar = c = k_B=1$.} 
If the system moves at a constant four-velocity $u^\mu$, this expression is easily generalized to\footnote{We use the metric convention $g^{\mu \nu}=\mbox{diag}(+1,-1,-1,-1)$.}
\begin{eqnarray}
\rho_{\text{geq} } = Z^{-1}_{\text{geq}} \exp\left(-\beta_\nu P^\nu \right)\;,\quad
Z_{\text{geq}} = \mathrm{Tr} \left[ \exp\left(-\beta_\nu P^\nu \right) \right] \;,\label{rho_geq}
\end{eqnarray}
where $\beta_\nu = \beta u_\nu$ and $P^\nu$ is the four-momentum operator.

For a system without a conserved charge which is in the hydrodynamic regime, instead of in global equilibrium, the relevant observable is the operator of the spacetime dependent energy-momentum tensor $T^{\mu \nu}(x)$. 
A reduced description can be achieved (and hydrodynamic equations can be derived) utilizing Zubarev's formalism of the nonequilibrium statistical operator \cite{Zubarev-1a,Zubarev-1b,Zubarev-1c,Zubarev-1d} (for modern developments, see Ref.~\cite{Zubarev-2} and references therein).
In Zubarev's approach \cite{Zubarev-1a} the nonequilibrium statistical operator maximizes the entropy subject to the following initial-state constraint imposed on a three-dimensional spacelike hypersurface $\sigma(\tau_0)$ with a timelike normal vector $n_{\mu}(x)$:
\begin{eqnarray}
n_{\mu} (x) \tilde{T}^{\mu \nu}(x) = n_{\mu} (x) \langle T^{\mu \nu}(x)\rangle\;. \label{4.4}
\end{eqnarray}
The left-hand side of the above equation is determined by a true state of the system, with energy-momentum tensor $\tilde{T}^{\mu \nu}(x)$, and the right-hand side features the expectation value of the energy-momentum tensor operator calculated with the nonequilibrium statistical operator, $\langle \cdots \rangle \equiv {\rm Tr}[\rho_{\text{neq}}[\sigma(\tau_0)] \cdots]$.
Maximizing the entropy is essentially synonymous to assuming that the initial state of the system is in local thermodynamical equilibrium.
Consequently, the nonequilibrium statistical operator reads \cite{Zubarev-3}
\begin{eqnarray}
\rho_{\text{neq}}[\sigma(\tau_0)]& =&  Z^{-1}_{\text{neq}}[\sigma(\tau_0)] \exp\left\{-\int_{\sigma(\tau_0) } \mathrm{d}\sigma_{\mu}(y)\, \beta_{\nu}(y) T^{\mu \nu}(y)\right\} \;, \nonumber \\ 
Z_{\text{neq}} [\sigma(\tau_0)]& = &  \mathrm{Tr} \left[ \exp \left\{ -\int_{\sigma(\tau_0) } \mathrm{d}\sigma_{\mu}(y)\, \beta_{\nu}(y) T^{\mu \nu}(y)\right\} \right]\;, \label{5}
\end{eqnarray}
where $\mathrm{d} \sigma_\mu(y) \equiv \mathrm{d} \sigma\, n_\mu(y)$. 
The inverse four-temperature vector $\beta_{\nu}(y)\equiv \beta(y) u_{\nu}(y)$ is now a spacetime dependent field, with $u_{\nu}(y)$ being the local four-velocity of the system (on a spacetime point $y$ on $\sigma(\tau_0)$), normalized as $u_\nu(y)u^{\nu}(y)=1$.
In this approach there is a continuum of constraints: on the whole hypersurface $\sigma(\tau_0)$ the relevant local observables, such as the energy-momentum tensor on the left-hand side of Eq.~\eqref{4.4}, are equal to the expectation values of the corresponding local quantum operators with respect to the statistical operator, i.e., the expectation value of the energy-momentum tensor operator on the right-hand side of Eq.~\eqref{4.4}.

The initial state, $\rho_{\text{neq}}[\sigma(\tau_0)]$, characterizes an actual state of the system.
This initial condition is not very restrictive if almost all initial states consistent with the above constraints evolve (in the Schr\"odinger picture) towards some hydrodynamical attractor (see, e.g., Ref.~\cite{Spalinski}), where such states become empirically indistinguishable with respect to the set of relevant observables and most details of the actual initial microscopic conditions become irrelevant.
In the original Zubarev approach the nonequilibrium statistical operator maximizes the entropy subject to constraints imposed in the infinitely remote past \cite{Zubarev-1a}.  
For the description of the transient evolution of matter in hadron and heavy-ion collisions it is more natural, however, to use initial conditions which correspond to the beginning of the collision, i.e., on a suitably chosen hypersurface $\sigma(\tau_0)$. 
The corresponding reformulation of Zubarev's method was discussed in Ref.~\cite{Zubarev-3}. 

In the Heisenberg picture, by definition the statistical operator does not change with time or, if some one-parameter family of three-dimensional spacelike hypersurfaces $\sigma(\tau)$ is defined, the statistical operator does not change with $\tau$. 
This allows us to take space-time gradients out of the statistical averaging with $\rho_{\text{neq}}[\sigma(\tau_0)]$, i.e., the conservation equations determining the Lagrange multipliers can be written as
\begin{eqnarray}
\langle \partial_{\mu} T^{\mu \nu}(x)\rangle=\partial_{\mu} \langle T^{\mu \nu}(x)\rangle = 0\;. \label{6}
\end{eqnarray}
Note that Eq.~(\ref{6}) is time reversible.\footnote{In the original Zubarev approach \cite{Zubarev-1a}, the reversibility of the exact conservation equations is broken by adding an infinitesimally small source term on the right-hand side of the evolution equation for the statistical operator in the Heisenberg representation.} 

Nevertheless, the fact that $\rho_{\text{neq}} [\sigma(\tau_{0})]$ does not change with time does not mean that it retains its functional form \eqref{5} on a three-dimensional spacelike hypersurface $\sigma(\tau)$ with $\tau > \tau_0$.
In fact, due to irreversible dissipative processes driven by spacetime gradients of the hydrodynamical variables (i.e, in our case $\beta_\nu$), the system will deviate from the initial local-equilibrium state on $\sigma(\tau_0)$ in the course of its evolution.
This can be seen as follows:
using Gauss' theorem and energy-momentum conservation one obtains
\begin{eqnarray}
- \int_{\sigma(\tau_0)} \mathrm{d} \sigma_{\mu}(y) \, \beta_\nu(y) T^{\mu\nu} (y)
   =  - \int_{\sigma(\tau)} \mathrm{d}  \sigma_{\mu}(y)  \, \beta_\nu(y) T^{\mu\nu}(y) +
 \int_{\Omega} \mathrm{d}^4z \, T^{\mu\nu}(z) \partial_{z,\mu} \beta_{\nu}(z)\;. \label{10}
\end{eqnarray}
Here, the four-dimensional spacetime volume $\Omega$ is enclosed by the two spacelike hypersurfaces $\sigma(\tau_0)$ and $\sigma(\tau)$ and timelike hypersurfaces connecting these two, where $\beta_{\nu}(y) T^{\mu\nu} (y) $ is supposed to vanish. 
Then, the nonequilibrium statistical operator can be written as \cite{Zubarev-3}
\begin{eqnarray}
\rho_{\text{neq}}[\sigma(\tau_0)]& =&  Z^{-1}_{\text{neq}}[\sigma(\tau_0)] \exp\left[-\int_{\sigma(\tau_0) } \mathrm{d}\sigma_{\mu}(y)\, \beta_{\nu}(y) T^{\mu \nu}(y)\right]  \nonumber \\ 
& = & Z^{-1}_{\text{neq}}[\sigma(\tau_0)] \exp\left[-\int_{\sigma(\tau) } \mathrm{d}\sigma_{\mu}(y)\, \beta_{\nu}(y) T^{\mu \nu}(y)+  \int_{\Omega} \mathrm{d}^4z \, T^{\mu\nu}(z) \partial_{z,\mu} \beta_{\nu}(z)\right] \;.
\end{eqnarray}
For the following, we define
\begin{eqnarray}
A & \equiv & \int_{\sigma(\tau)} \mathrm{d} \sigma_{\mu}(y)\,  \beta_\nu(y) T^{\mu\nu}(y)\;, \\
B & \equiv & \int_{\Omega} \mathrm{d}^4z\,  T^{\mu\nu} (z)\partial_{z,\mu} \beta_{\nu}(z)\;,
\end{eqnarray}
and assume that $B$ is small compared to $A$, which is true if $\partial_{z,\mu} \beta_{\nu}(z)$ is sufficiently small. 
Consequently, to leading (zeroth) order in $B$ the nonequilibrium statistical operator is equal to the \textit{local-equilibrium} statistical operator on the hypersurface $\sigma(\tau)$ \cite{Zubarev-3}
\begin{eqnarray}
\rho_{\text{leq}}[\sigma(\tau)]& =& Z^{-1}_{\text{leq}}[\sigma(\tau)] \exp\left[-\int_{\sigma (\tau) } \mathrm{d}\sigma_{\mu}(y)\,\beta_{\nu}(y) T^{\mu \nu}(y)\right]\; , \nonumber \\
Z_{\text{leq}}[\sigma(\tau)] & = & \mathrm{Tr} \left[  \exp\left\{-\int_{\sigma (\tau) } \mathrm{d}\sigma_{\mu}(y)\,\beta_{\nu}(y) T^{\mu \nu}(y)\right\}\right]\;.\label{11}
\end{eqnarray}
Corrections to the leading order can be computed perturbatively; to linear (first) order in $B$, one obtains \cite{Zubarev-3}
\begin{eqnarray}
\rho_{\text{neq}}[\sigma(\tau_0)]& =& \rho_{\text{leq}}[\sigma(\tau)] \left( 1 + \int_0^1 \mathrm{d}\lambda\, e^{\lambda A} B e^{-\lambda A} - \langle B \rangle_{\text{leq}} \right) + \mathcal{O}(B^2)\;, \label{linear_in_B}
\end{eqnarray}
where $\langle \cdots \rangle_{\text{leq}} \equiv \mathrm{Tr}\left[ \rho_{\text{leq}} [\sigma(\tau)] \cdots \right]$.
Since $B \sim T^{\mu \nu}$, calculating the expectation value of the energy-momentum tensor, $\langle T^{\mu \nu}(x) \rangle$, including these corrections one obtains terms involving two-point correlation functions of the energy-momentum tensor.
These can be expressed in terms of transport coefficients using the well-known Kubo relations.
One thus obtains the well-known dissipative terms in the equations of motion of dissipative hydrodynamics. 
Since the transport coefficients are proportional to the mean free path $\lambda_{\text{mfp}}$ of particles in the system, while $\partial \beta$ is inversely proportional to the length of homogeneity $L$ of the system, the dissipative terms are proportional to the Knudsen number $\mathrm{Kn} \equiv \lambda_{\text{mfp}}/L$.

The transport coefficients are calculated in the Markovian (short-memory) approximation, which exploits the existence of disparate time scales in the system. 
It is worth noting that this approximation is not only a useful tool for explicit calculations: the Markovian level of description mimics the effective loss of details about the initial conditions. 
Irreversible dissipative hydrodynamics appears as an effective theory of the slow degrees of freedom in its range of applicability, see, e.g., Refs.~\cite{Zubarev-1a,Zubarev-1b,Zubarev-1c,Zubarev-1d,Zubarev-2,Zubarev-3} for derivations of dissipative hydrodynamical equations in the Zubarev approach.

Note that if one neglects the dissipative corrections, i.e., sets $B=0$, the statistical operator \eqref{11} actually retains its initial form \eqref{5}, i.e., local thermodynamical equilibrium is maintained throughout the evolution of the system.
One would now naively assume that, if one calculates $\langle T^{\mu \nu}(x) \rangle$ by setting $B=0$ in Eq.~\eqref{linear_in_B}, i.e., if one calculates $\langle T^{\mu \nu} (x) \rangle_{\text{leq}}$, one would obtain the ideal-fluid form for the energy-momentum tensor in local thermodynamical equilibrium,
\begin{equation}
\tilde{T}^{\mu \nu}_{\text{id}}(x)  = \left[\epsilon(x) + \mathcal{P}(x)\right] u^\mu(x) u^\nu(x) - \mathcal{P}(x) g^{\mu \nu}\;,
\label{idealfluid}
\end{equation}  
where $\epsilon(x)$ and $\mathcal{P}(x)$ are the local energy density and pressure of the fluid, respectively.
However, this is not true, in fact $\langle T^{\mu \nu} (x) \rangle_{\text{leq}} \neq  \tilde{T}^{\mu \nu}_{\text{id}}(x)$, due to the nonlocality of $\rho_{\text{leq}}[\sigma(\tau)]$ introduced by the integration over the hypersurface $\sigma(\tau)$.
In order to obtain the ideal-fluid form, one requires an additional approximation. 
Namely, one needs to assume that $\beta_{\mu}(y)$ varies sufficiently smoothly over the hypersurface $\sigma(\tau)$, such that the expectation value of some local operator $O(x)$, $\langle O(x) \rangle_{\text{leq}} = {\rm Tr}[\rho_{\text{leq}} [\sigma(\tau)] O(x)]$, will be mainly determined by the value of the field $\beta_{\mu}(y)$ around the point $x$.  
Corrections can be systematically taken into account by expanding $\beta_{\mu}(y)$ in a Taylor series around $x$,
\begin{eqnarray}
\beta_{\mu}(y)=\beta_{\mu}(x) + \partial_{\lambda}\beta_{\mu}(x)(y-x)^{\lambda} + \mathcal{O}\left((\partial \beta)^2\right)\;,
 \label{12}
\end{eqnarray}
and then substituting it into $\rho_{\text{leq}}[\sigma(\tau)]$. 
Introducing the four-momentum operator of the system on the hypersurface $\sigma(\tau)$, $P^\nu \equiv \int_{\sigma(\tau)}\mathrm{d}\sigma\, n_{\mu}(y) T^{\mu \nu}(y)$, and denoting
\begin{equation}
C \equiv \partial_\lambda \beta_\nu(x) \int_{\sigma(\tau)} \mathrm{d} \sigma_\mu(y)\, (y-x)^\lambda T^{\mu \nu}(y)
+  \mathcal{O}\left((\partial \beta)^2\right)\;,
\end{equation}
we then obtain
\begin{equation}
\int_{\sigma (\tau) } \mathrm{d}\sigma_{\mu}(y)\,\beta_{\nu}(y) T^{\mu \nu}(y)
= \beta_\nu(x) P^\nu + C\;. \label{Taylorbeta}
\end{equation}
If $C$ is a small correction to $\beta_\nu(x) P^\nu$, expectation values with respect to $\rho_{\text{leq}}[\sigma(\tau)]$ can be calculated perturbatively in a power series in $C$, just as the one in $B$ in Eq.~\eqref{linear_in_B} (see, e.g., Ref.~\cite{Becattini-1}). 
Since $C \sim T^{\mu \nu}$, to linear order the expectation value of the energy-momentum tensor receives corrections proportional to the two-point function of the energy-momentum tensor.
In contrast to the terms $\sim B$ discussed previously, however, there are no traditional Kubo relations to relate these corrections to transport coefficients, since now there is no spacetime integral over $\Omega$, but only an integral over the hypersurface $\sigma(\tau)$.

To leading (zeroth) order, we set $C=0$ and insert Eq.~\eqref{Taylorbeta} into $\rho_{\text{leq}}[\sigma(\tau)]$. 
One then obtains an expression which is formally identical to the global-equilibrium statistical operator \eqref{rho_geq}, but with a \textit{spacetime dependent} $\beta_\nu(x)$, which we refer to as ``spacetime ($x$-)dependent global equilibrium'' ($x$-geq),
\begin{eqnarray}
\rho_{\text{leq}}[\sigma(\tau)] \;\; \stackrel{C=0}{\longrightarrow} \;\; \rho_{x-\text{geq} }(x) & = & Z^{-1}_{x-\text{qeq}}(x) \exp\left[-\beta_{\nu}(x)P^{\nu}\right]\;, \nonumber \\
Z_{x-\text{geq}}(x) & = & \mathrm{Tr} \left[ \exp\left\{-\beta_{\nu}(x)P^{\nu}\right\}\right]\;.
 \label{4.1}
\end{eqnarray}
It is now obvious (by reasons of symmetry alone) that
\begin{equation}
\langle T^{\mu \nu}(x) \rangle_{x-\text{geq}} = \mathrm{Tr} \left[ \rho_{x-\text{geq}} T^{\mu \nu}(x) \right]
= \tilde{T}^{\mu \nu}_{\text{id}}(x)\;. \label{Tmunu_ideal_2}
\end{equation}
For this reason, the spacetime ($x$-)dependent global-equilibrium energy-momentum tensor is usually referred to as energy-momentum tensor in ``local equilibrium''. 
We have seen that this is not quite correct, as the true local-equilibrium energy-momentum tensor, $\langle T^{\mu \nu} (x) \rangle_{\text{leq}}$, contains additional gradient terms $\sim C$ (and powers thereof). 
Therefore, although using ``spacetime dependent global equilibrium'' sounds contradictory and much more cumbersome than ``local equilibrium'', we will stick to this nomenclature in the following, since it accurately expresses the fact that expectation values are computed with the operator~\eqref{4.1} instead of the local-equilibrium statistical operator~\eqref{11}.

In this paper, we calculate the expectation value of the energy-momentum tensor operator of a real scalar field $\phi(x)$, using similar approximations for the nonequilibrium statistical operator as outlined above. 
However, we deviate from the above approach, which yields the well-known result \eqref{Tmunu_ideal_2}, in one important aspect.
Namely, the explicit expression for the energy-momentum tensor operator in general involves spacetime gradients of the field operators. 
Using the stationarity of the nonequilibrium statistical operator \eqref{5}, we can take these derivatives outside of the statistical averaging,  as in Eq.~\eqref{6}. 
The approximations outlined above are then applied to the calculation of the statistical two-point function of the fields, 
$\langle \phi(x) \phi(y) \rangle$, rather than to the calculation of the expectation value $\langle T^{\mu \nu}(x) \rangle$ of the full energy-momentum tensor operator.
We show that the expression obtained for $\langle T^{\mu \nu}(x) \rangle$ under these approximations involves corrections with respect to the ideal-fluid form \eqref{idealfluid}.\footnote{See also Refs. \cite{local,local-2}.} 
In particular, we demonstrate that the relation between pressure and energy density, i.e., the equation of state, is affected by these corrections and therefore is modified with respect to the case of global thermodynamical equilibrium.
The magnitude of these corrections is determined by the ratio of the thermal wavelength $\lambda_{\text{th}}$ to the typical spacetime homogeneity length $L$ of a given system. 
If the latter is much larger than the former, then these additional terms can be neglected.  
This is certainly true for the particular limit of global thermodynamical equilibrium, as well as for a coarse-grained description of  macroscopic systems, where there is a clear separation between the microscopic and the macroscopic scales. 
However, this need not be the case, for example, for relativistic hadron and heavy-ion collisions, where small and highly inhomogeneous systems are created with a typical macroscopic time or length scale of a few femtometers.

The remainder of this paper is structured as follows: 
in Sec.~\ref{Tmunu} we compute the expectation value of the energy-momentum tensor based on the idea outlined above, using a real scalar field as a simple, yet explicit example. 
In Sec.~\ref{corrections} we derive the corrections to the equation of state in global thermodynamical equilibrium arising from spacetime gradients of the hydrodynamical fields. 
Our conclusions are given in Sec.~\ref{conclusions}.

\section{Alternative calculation of the expectation value of the energy-momentum tensor}
\label{Tmunu}

For the sake of simplicity, we consider a real scalar field with the action
\begin{eqnarray}
S=\int \mathrm{d}^{4}x \, \mathcal{L}\; ,
\label{4.2}
\end{eqnarray}
with $\mathcal{L}$ being the corresponding Lagrangian
\begin{eqnarray}
\mathcal{L}=\frac{1}{2} \partial_{\mu}\phi\partial^{\mu}\phi
- \frac{m^{2}}{2}\phi^{2} + \mathcal{L}_{\text{int}}(\phi)\;,
\label{4.3}
\end{eqnarray}
where the interacting part of the Lagrangian density, $\mathcal{L}_{\text{int}}(\phi)$, does not contain spacetime derivatives. 
To calculate $\langle T^{\mu \nu}(x)\rangle$, one needs an explicit form for the operator of the energy-momentum tensor. 
The canonical one reads \begin{eqnarray}
T^{\mu \nu}(x)=\partial^{\mu}\phi\partial^{\nu}\phi -g^{\mu \nu}\mathcal{L}\;,
\label{8}
\end{eqnarray}
where the Lagrangian $\mathcal{L}$ is given by Eq.~(\ref{4.2}).
Following the idea formulated in the introduction, one can take the spacetime derivatives in Eq.~(\ref{8}) outside the statistical averaging, just as in Eq.~(\ref{6}), since the nonequilibrium statistical operator is stationary.
We then obtain\footnote{For simplicity, we will assume that the field expectation value vanishes, $\langle \phi (x) \rangle =0$.}
\begin{eqnarray}
\langle T^{\mu \nu}(z) \rangle =
\left(\partial^{\mu}_{x}\partial^{\nu}_{y}-\frac{1}{2}g^{\mu\nu} \partial_{x\alpha}\partial^{\alpha}_{y}\right)
F(x,y)\Big|_{{x=y = z}} +  g^{\mu \nu}\left(\frac{1}{2}m^{2}\langle \phi^{2}(z)\rangle -\langle \mathcal{L}_{\text{int}}(\phi)\rangle \right)\;.
\label{9}
\end{eqnarray}
Here $F(x,y)$ denotes the statistical two-point function,
\begin{eqnarray}
F(x,y)=\langle \hat{F}(x,y) \rangle\;, \label{9.01} \\
\hat{F} (x,y)  = \frac{1}{2}\{\phi(x),\phi(y)\} \equiv \frac{1}{2}\left[\phi(x)\phi(y)+\phi(y)\phi(x)\right] \; .
\label{9.02}
\end{eqnarray}
It is convenient to introduce the new variables
\begin{eqnarray}
Z = \frac{x+y}{2}\;, \label{9.1} \\
\Delta z = y-x\;.
\label{9.2}
\end{eqnarray}
Then,
\begin{eqnarray}
\langle T^{\mu \nu}(z) \rangle & = &
\left(-\partial^{\mu}_{\Delta z}\partial^{\nu}_{\Delta z}+\frac{1}{2}g^{\mu\nu} \partial_{\Delta z,\alpha}\partial^{\alpha}_{\Delta z}\right)
F\left(Z -\frac{\Delta z}{2},Z+\frac{\Delta z}{2} \right)\Big|_{{Z = z,\Delta z=0}}  \nonumber \\ & &+ \left(\frac{1}{4}\partial^{\mu}_{Z}\partial^{\nu}_{ Z}
-\frac{1}{8} g^{\mu\nu}\partial_{ Z,\alpha}\partial^{\alpha}_{ Z} \right)F\left(Z -\frac{\Delta z}{2},Z+\frac{\Delta z}{2} \right)\Big|_{{Z = z,\Delta z=0}}
 \nonumber \\ & &+ \, g^{\mu\nu}\left(\frac{1}{2}m^{2}\langle \phi^{2}(z )\rangle-\langle \mathcal{L}_{\text{int}}(\phi)\rangle\right)\;.
\label{9.21}
\end{eqnarray}
The result can be compactly written as
\begin{eqnarray}
\langle T^{\mu \nu}(z) \rangle & = & \int \mathrm{d}^{4}\Delta z\,  \delta^{(4)}(\Delta z)
\left(-\partial^{\mu}_{\Delta z}\partial^{\nu}_{\Delta z}+\frac{1}{2}g^{\mu\nu} \partial_{\Delta z,\alpha}\partial^{\alpha}_{\Delta z}\right)
F\left(z -\frac{\Delta z}{2},z +\frac{\Delta z}{2} \right)  \nonumber \\ & & + \left(\frac{1}{4}\partial^{\mu}_{z}\partial^{\nu}_{ z}
-\frac{1}{8} g^{\mu\nu}\partial_{ z,\alpha}\partial^{\alpha}_{ z} \right)\langle \phi^{2}(z )\rangle
+g^{\mu\nu}\left(\frac{1}{2}m^{2}\langle \phi^{2}(z )\rangle-\langle \mathcal{L}_{\text{int}}(\phi)\rangle\right)\;.
\label{9.3}
\end{eqnarray}
As advertised in the introduction, we now replace the statistical average with respect to the nonequilibrium statistical operator, $\langle \cdots \rangle = \mathrm{Tr}[\rho_{\text{neq}}[\sigma(\tau_0)] \cdots ]$, by the thermal average with respect to the spacetime dependent global-equilibrium operator at point $z$, $\langle \cdots \rangle_{z-\text{geq}} = \mathrm{Tr}[\rho_{z-\text{geq}}(z) \cdots ] $, i.e., we neglect all dissipative corrections $\sim B, C$.
We will call the result the ``gradient-corrected $z$-dependent global-equilibrium approximation'' (gc-$z$-geq), which is then
\begin{align}
\tilde{T}^{\mu \nu}_{\text{gc-$z$-geq}}(z)  & =  \int \mathrm{d}^{4}\Delta z \, \delta^{(4)}(\Delta z)
\left(-\partial^{\mu}_{\Delta z}\partial^{\nu}_{\Delta z} + \frac{1}{2}g^{\mu\nu} \partial_{\Delta z,\alpha}\partial^{\alpha}_{\Delta z}\right)
F_{z-\text{geq}}\left(z -\frac{\Delta z}{2},z +\frac{\Delta z}{2} \right)  \nonumber \\ 
 +  & \left(\frac{1}{4}\partial^{\mu}_{z}\partial^{\nu}_{ z}
-\frac{1}{8} g^{\mu\nu}\partial_{ z,\alpha}\partial^{\alpha}_{ z} \right)\langle \phi^{2}(z )\rangle_{z-\text{geq}}
+g^{\mu\nu}\left(\frac{1}{2}m^{2}\langle \phi^{2}(z )\rangle_{z-\text{geq}}-\langle \mathcal{L}_{\text{int}}(z)\rangle_{z-\text{geq}}\right)\;,
\label{13}
\end{align}
where $F_{z-\text{geq}}\left(z -\frac{\Delta z}{2},z +\frac{\Delta z}{2} \right)=\langle \hat{F}\left(z -\frac{\Delta z}{2},z +\frac{\Delta z}{2} \right) \rangle_{z-\text{geq}}$. 
Equation (\ref{13}) expresses the expectation value of the energy-momentum tensor in terms of the thermal $n$-point functions and their derivatives.
Regularization of the above expression can be done by subtracting the corresponding expectation values at zero temperature from the thermal $n$-point functions. 

The terms in Eq.~(\ref{13}) with derivatives with respect to $z$ manifest deviations from (spacetime dependent) global thermodynamical equilibrium where only relative positions matter.
Therefore it is convenient to write the result as 
\begin{eqnarray}
\tilde{T}^{\mu \nu}_{\text{gc-$z$-geq}}(z) = \tilde{T}^{\mu \nu}_{z-\text{geq}}(z) + \left(\frac{1}{4}\partial_z^{\mu}\partial_z^{\nu}
-\frac{1}{8} g^{\mu\nu}\partial_{ z,\alpha}\partial_z^{\alpha} \right)\langle \phi^{2}(z )\rangle_{z-\text{geq}}\;, \label{15}
\end{eqnarray}
where 
\begin{eqnarray}
 \tilde{T}^{\mu \nu}_{z-\text{geq}}(z) & = &\int \mathrm{d}^{4}\Delta z \, \delta^{(4)}(\Delta z)
\left(-\partial^{\mu}_{\Delta z}\partial^{\nu}_{\Delta z} + \frac{1}{2}g^{\mu\nu} \partial_{\Delta z,\alpha}\partial^{\alpha}_{\Delta z}\right)
F_{z-\text{geq}}\left(z -\frac{\Delta z}{2},z +\frac{\Delta z}{2} \right)  \nonumber \\ &  & + g^{\mu\nu}\left(\frac{1}{2}m^{2}\langle \phi^{2}(z )\rangle_{z-\text{geq}}-\langle \mathcal{L}_{\text{int}}(z)\rangle_{z-\text{geq}}\right)\;.
\label{16}
\end{eqnarray}
As we will now show, $\tilde{T}^{\mu \nu}_{z-\text{geq}}(z)$ coincides with the expectation value of the energy-momentum tensor calculated with the $z$-dependent global-equilibrium statistical operator \eqref{4.1}, i.e., we will show that $\tilde{T}^{\mu \nu}_{z-\text{geq}}(z) = \tilde{T}^{\mu \nu}_{\text{id}}(z)$, in agreement with Eq.~\eqref{Tmunu_ideal_2}. 
To this end, let us Fourier-transform the statistical two-point function  $F_{z-\text{geq}}\left(z -\frac{\Delta z}{2},z +\frac{\Delta z}{2} \right)$ with respect to the relative coordinate $\Delta z$, as follows:
\begin{eqnarray}
F_{z-\text{geq}}\left(z -\frac{\Delta z}{2},z +\frac{\Delta z}{2} \right) = \frac{1}{(2\pi)^{4}}\int \mathrm{d}^{4}p\, e^{i\Delta z_\mu \,p^\mu}G_{z-\text{geq}} (p,z)\;. \label{20.3}
\end{eqnarray}
Substituting Eq.~(\ref{20.3}) into Eq.~(\ref{16}) we obtain
\begin{eqnarray}
\tilde{T}^{\mu \nu}_{z-\text{geq}}(z) & = &\frac{1}{(2\pi)^{4}}\int \mathrm{d}^{4}p \, p^{\mu}p^{\nu}G_{z-\text{geq}} (p,z)
 \nonumber \\ & + & \frac{g^{\mu \nu}}{2(2\pi)^{4}}\int \mathrm{d}^{4}p \, (m^{2}-p^{2})G_{z-\text{geq}} (p,z) - g^{\mu \nu}\langle \mathcal{L}_{\text{int}}(\phi)\rangle_{z-\text{geq}}\; . \label{21}
\end{eqnarray}
In the following, for the sake of simplicity we set the interactions to zero, $\mathcal{L}_{\text{int}} (\phi) =0$, and use the kinetic on-mass-shell  approximation (see, e.g., Ref.~\cite{Groot}), where $G_{z-\text{geq}} (p,z)= 4 \pi \delta(p^2-m^2) \theta(p_0) n_B(\beta_{\mu}(z)p^{\mu})$, with $n_B(x) = (e^{x}-1)^{-1}$ being the Bose-Einstein distribution function.
Then, the last two terms in Eq.~\eqref{20.3} vanish.
We arrive at\footnote{For the sake of simplicity we suppress the dependence on $z$.}
\begin{eqnarray}
\tilde{T}^{\mu \nu}_{z-\text{geq}} = \frac{1}{(2\pi)^{4}}\int \mathrm{d}^{4}p \, p^{\mu}p^{\nu}G_{z-\text{geq}}(\beta_{\mu}p^{\mu}) = (\epsilon_{z-\text{geq}} + {\cal P}_{z-\text{geq}})u^{\mu}u^{\nu} - {\cal P}_{z-\text{geq}}g^{\mu\nu}  \;.
\label{22}
\end{eqnarray}
Here,
\begin{eqnarray}
\epsilon_{z-\text{geq}} & = & \frac{1}{(2\pi)^{3}}\int \frac{\mathrm{d}^{3}\mathbf{p}}{p_0} \, u_{\mu} p^{\mu} u_{\nu}p^{\nu}n_B(\beta_{\mu}p^{\mu}) \; , \label{23} \\
{\cal P}_{z-\text{geq}}& = & -\frac{1}{3}\Delta_{\mu \nu}\frac{1}{(2\pi)^{3}}\int \frac{\mathrm{d}^{3}\mathbf{p}}{p_0} \, p^{\mu}p^{\nu} n_B(\beta_{\mu}p^{\mu})\;,
\label{24}
\end{eqnarray}
where $p_0 = \sqrt{\mathbf{p}^2 + m^2}$ is the on shell energy of the particles and $\Delta_{\mu \nu}\equiv g_{\mu\nu}-u_{\mu}u_{\nu}$ is the projector onto the three-dimensional subspace orthogonal to $u^\mu$.
Equations \eqref{23} and \eqref{24} are the standard expressions for the energy density and pressure of a 
noninteracting Bose gas of particles with mass $m$.
We thus confirm that $\tilde{T}^{\mu \nu}_{z-\text{geq}}(z) = \tilde{T}^{\mu \nu}_{\text{id}}(z)$.

Let us conclude this section by mentioning that the other terms on the right-hand side of Eq.~(\ref{15}) represent gradient corrections to the energy-momentum tensor of an ideal fluid in $z$-dependent global equilibrium.
Usually, it is assumed that $\tilde{T}^{\mu \nu}_{\text{gc-$z$-geq}}(z) \simeq \tilde{T}^{\mu \nu}_{z-\text{geq}}(z) = \tilde{T}^{\mu \nu}_{\text{id}}(z)$.
This approximation is certainly true for isotropic systems close to thermodynamical equilibrium, but can be questioned for small and highly inhomogeneous systems.
In the next section, we will explicitly compute these gradient corrections.

\section{Pseudogauge dependence and gradient-corrected spacetime dependent energy-momentum tensor}
\label{corrections}

Equation (\ref{15}) can be written in a more suggestive way as follows: 
\begin{eqnarray}
\tilde{T}^{\mu \nu}_{\text{gc-$z$-geq}}(z) =\tilde{T}^{\mu \nu}_{z-\text{geq}}(z) + \Lambda (z)g^{\mu\nu} +\Phi^{\mu \nu} (z)\;, \label{17}
\end{eqnarray}
where 
\begin{eqnarray}
\Lambda (z) = \frac{1}{8} \partial_{z, \alpha}\partial_z^{\alpha} \langle \phi^{2}(z )\rangle_{z-\text{geq}}\;, \label{18} \\
\Phi^{\mu \nu} (z) = \frac{1}{4}(\partial_z^{\mu}\partial_z^{\nu}
- g^{\mu\nu}\partial_{z,\alpha}\partial_z^{\alpha} )\langle \phi^{2}(z )\rangle_{z-\text{geq}}\;. \label{19}
\end{eqnarray}
One observes that 
\begin{eqnarray}
\partial_{z,\mu}\Phi^{\mu \nu} (z) = 0\;, \label{20}
\end{eqnarray}
and, therefore, $\Phi^{\mu \nu}$ does not contribute to the energy-momentum conservation equation. 
In effect, $\Phi^{\mu \nu}$ influences the initial conditions only. 

A natural question to ask is whether one can redefine the energy-momentum tensor in such a way that $\Phi^{\mu \nu} \equiv 0$. 
One can show that this can be done by means of modifying the energy-momentum tensor (\ref{8}) by a so-called \textit{pseudogauge transformation}, i.e., in essence by adding suitable terms which do not affect the energy-momentum conservation equation,\footnote{It has been known for a long time that there exists an ambiguity in the definition of the energy-momentum tensor. The energy-momentum tensor is known to be defined only up to the derivative of an antisymmetric tensor. The possibility of a simultaneous modification of both the energy-momentum and spin tensors was discussed in Ref.~\cite{Hehl}.}
\begin{eqnarray}
\mathcal{T}^{\mu\nu} =  T^{\mu\nu} - \frac{1}{4}(\partial^{\mu}\partial^{\nu}
- g^{\mu\nu}\partial_{ \alpha}\partial^{\alpha} ) \phi^{2}\;, \label{20.1}
\end{eqnarray}
to obtain the desired result.\footnote{This energy-momentum tensor can 
be  also obtained by adding a total derivative to the Lagrangian,
$\mathcal{L} \rightarrow \mathcal{L} -\frac{1}{2}\partial_{\mu}(\phi \partial^{\mu}\phi)$. } 
Note that this energy-momentum tensor yields the same four-momentum as the canonical one, and  $\partial_{\mu}{\cal T}^{\mu \nu}  = 0$.
Equation \eqref{17} implies that
\begin{eqnarray}
\tilde{\mathcal{T}}^{\mu \nu}_{\text{gc-$z$-geq}}(z) =\tilde{T}^{\mu \nu}_{z-\text{geq}}(z) + \Lambda (z)g^{\mu\nu} \;. \label{20.2}
\end{eqnarray}

Substituting Eq.~(\ref{22}) into Eq.~(\ref{20.2}) we readily see that $\tilde{\mathcal{T}}^{\mu \nu}_{\text{gc-$z$-geq}}$ has the ideal-fluid form,
\begin{eqnarray}
\tilde{\mathcal{T}}^{\mu \nu}_{\text{gc-$z$-geq}}& =& (\epsilon_{z-\text{geq}} + {\cal P}_{z-\text{geq}})u^{\mu}u^{\nu} - {\cal P}_{z-\text{geq}}g^{\mu\nu} + \Lambda g^{\mu\nu} \label{25.1} \\
& = & (\epsilon_{\text{gc-$z$-geq}} + {\cal P}_{\text{gc-$z$-geq}})u^{\mu}u^{\nu} - {\cal P}_{\text{gc-$z$-geq}}g^{\mu\nu}\;,  \label{26}
\end{eqnarray}
where 
\begin{eqnarray}
\epsilon_{\text{gc-$z$-geq}} & = & \epsilon_{z-\text{geq}}  + \Lambda\;,  \label{27} \\
{\cal P}_{\text{gc-$z$-geq}} & = & {\cal P}_{z-\text{geq}} - \Lambda \;, \label{28}\\
\Lambda & =& \frac{1}{8}\partial_{ \alpha}\partial^{\alpha} \frac{1}{(2\pi)^{3}}\int \frac{\mathrm{d}^{3}\mathbf{p}}{p_0}\, n_B(\beta_{\mu}p^{\mu})\;.
\label{29}
\end{eqnarray}
One observes that $\epsilon_{\text{gc-$z$-geq}}$ is the (spacetime dependent) energy density and $u^{\mu}$ is the hydrodynamical four-velocity in the Landau frame of $\tilde{\mathcal{T}}^{\mu \nu}_{\text{gc-$z$-geq}}(z)$. 
Using Eqs.~(\ref{26}), (\ref{27}), (\ref{28}), and (\ref{29}) we conclude that $\tilde{\mathcal{T}}^{\mu \nu}_{\text{gc-$z$-geq}}(z)$ leads to the differential equations of ideal hydrodynamics with an equation of state, which, for $\Lambda \neq 0$, is not equal to that in (spacetime dependent) global thermodynamical equilibrium.

The expressions for the energy density and pressure obtained in Eqs.~(\ref{27}) and (\ref{28}) contain corrections with respect to the global-equilibrium case. 
In effect, these contributions to the energy-momentum tensor are negligibly small if 
\begin{eqnarray}
\frac{|\Lambda |}{{\cal P}_{z-\text{geq}}}\ll 1\;,  \label{30}
\end{eqnarray}
see Eq.~(\ref{25.1}). 
To estimate the magnitude of these corrections, one would in principle first have to solve the hydrodynamic equations with $\Lambda =0$ for $\beta$ and $u^{\mu}$, and then use these values for $\beta$ and $u^{\mu}$ to evaluate $|\Lambda |/{\cal P}_{z-\text{geq}}$. 
To get an intuitive understanding of the magnitude of  $|\Lambda |/{\cal P}_{z-\text{geq}}$, let us compute the integral in Eq.~(\ref{24}) in the local rest frame of the fluid, where $u^{* \mu }=(1,0,0,0)$ and $u^{*}_\mu p^{*\mu}= p_0^{*} = \sqrt{\mathbf{p}^{* 2} + m^2}$.
The result is 
\begin{eqnarray}
{\cal P}_{z-\text{geq}}& =& \frac{1}{3}\frac{1}{(2\pi)^{3}}\int \frac{\mathrm{d}^{3}\mathbf{p}^{*}}{p_0^*}\, \mathbf{p}^{*2}
n_B (\beta p_0^{*}) \;.
\label{31}
\end{eqnarray}
It is convenient  to introduce the following notations: 
\begin{eqnarray}
\overline{\mathbf{p}^{*2}}& \equiv & \frac{\int \frac{\mathrm{d}^{3}\mathbf{p}^{*}}{p_0^*}\, \mathbf{p}^{*2}
n_B (\beta p_0^{*})}{\int \frac{\mathrm{d}^{3}\mathbf{p}^{*}}{p_0^*}\, 
n_B (\beta p_0^{*})}\;,
\label{40} \\
\frac{1}{L^{2}}& \equiv & \frac{|\partial_{\alpha}\partial^{\alpha} \int \frac{\mathrm{d}^{3}\mathbf{p}^{*}}{p_0^*}\,
n_B (\beta p_0^{*})|}{\int \frac{\mathrm{d}^{3}\mathbf{p}^{*}}{p_0^*}\, n_B (\beta p_0^{*})}\;.
\label{41}
\end{eqnarray}
Then 
\begin{eqnarray}
{\cal P}_{z-\text{geq}}& = & \frac{1}{3}\overline{\mathbf{p}^{*2}}\int \frac{\mathrm{d}^{3}\mathbf{p}^{*}}{p_0^*}\, n_B (\beta p_0^{*})\; ,
\label{41.1}\\
|\Lambda| & =& \frac{1}{8}\frac{1}{L^{2}}\int \frac{\mathrm{d}^{3}\mathbf{p}^{*}}{p_0^*}\, \mathbf{p}^{*2}
n_B (\beta p_0^{*})\;,
\label{41.2}
\end{eqnarray}
and $|\Lambda |/{\cal P}_{z-\text{geq}}$ reads
\begin{eqnarray}
\frac{|\Lambda |}{{\cal P}_{z-\text{geq}}}=\frac{3}{8}\frac{1}{\overline{\mathbf{p}^{*2}}L^{2}}\;. \label{42}
\end{eqnarray}
For the purpose of illustration, let us consider the nonrelativistic, $m\beta =m/T \gg 1$, and ultrarelativistic, $m\beta =m/T \ll 1$, limits.
Equation (\ref{42}) then becomes 
\begin{eqnarray}
\frac{|\Lambda |}{{\cal P}_{z-\text{geq}}}  \sim \frac{\lambda^{2}_{\text{th}}}{L^{2}} \;, \label{43}
\end{eqnarray}
where $\lambda_{\text{th}} $ is the thermal wavelength: $\lambda_{\text{th}} \sim 1/\sqrt{mT}$ for $m\beta \gg 1$ and $\lambda_{\text{th}} \sim 1/T$ for $m\beta \ll 1$.
Meanwhile, $L$ can be interpreted as the spacetime length of homogeneity of a system, see Eq.~(\ref{41}).
We thus conclude that the  corrections to the global-equilibrium equation of state can be neglected if the thermal wavelengths are much smaller than the characteristic spacetime scales of the system, 
\begin{eqnarray}
 \frac{\lambda_{\text{th}}}{L} \ll 1\; . \label{44}
\end{eqnarray}
Otherwise, if $\lambda_{\text{th}}/L \gtrsim 1$,  then the  energy density,  $\epsilon_{z-\text{leq}}$, or pressure, ${\cal P}_{z-\text{leq}}$, can become negative, see Eqs.~(\ref{27}), (\ref{28}), and (\ref{29}). 
Such a violation of the positivity condition implies the inapplicability of hydrodynamics,\footnote{See also Ref.~\cite{small}, where the domain of validity of hydrodynamics applied to small systems was considered.} 
and suggests that spacetime gradients in the system are so large that one can neither neglect the standard dissipative corrections, nor those arising from spacetime gradients of the ideal hydrodynamical variables such as $\beta_\nu$.
It is worth noting that the inequality  (\ref{44}) is reminiscent of the condition that the Knudsen number should be  sufficiently small for the applicability of the hydrodynamical description. 
Then, our analysis supports the view that, as an effective theory, hydrodynamics is applicable if there is a clear separation between the microscopic and the macroscopic scales. 

As a final comment, we note that if one utilizes the Klein-Gordon Lagrangian for spinors built with second-order derivatives of the fields (for a review of the most common choices of the energy-momentum and spin tensors for Dirac fields, see, e.g., Ref.~\cite{Speranza} ), then the considered corrections should also manifest itself for the Dirac field. 
However, this issue is beyond the scope of this study.

\section{Conclusions}
\label{conclusions}

In the present paper, we calculated nondissipative corrections to the energy-momentum tensor of a real scalar field in a state which is described by a statistical operator with a spacetime dependent temperature four-vector and which is commonly identified with a local-equilibrium state. 
In particular, we demonstrated that the relation between pressure and energy density, i.e., the equation of state, is affected by these corrections and therefore is modified with respect to the case of global thermodynamical equilibrium. 
The corrections are of second order in the ratio of the thermal wavelength to the typical macroscopic
length scale of the system.
If the thermal wavelengths are comparable in size with the typical spatiotemporal scales of the system, then a hydrodynamical description might fail. 
Our findings support the conjecture that hydrodynamics is based on a clear separation of scales: the microscopic scales, which are characteristic of the underlying microscopic theory, e.g., a quantum field theory, and the macroscopic scales, which characterize the spatiotemporal variation of the hydrodynamical variables. 
Our results present a challenge for the hydrodynamical description of the small, inhomogeneous systems created in hadron or hadron-ion collisions.

\begin{acknowledgments}
The authors thank F.~Becattini for clarifying remarks and discussions.
S.V.A.~acknowledges support by a grant from the Simons Foundation
(Grant No. 1290596).
D.H.R.~acknowledges support by the Deutsche Forschungsgemeinschaft (DFG, German Research Foundation) through the CRC-TR 211 ``Strong-interaction matter under extreme conditions'' -- Project No. 315477589 -- TRR 211, and by the State of Hesse within the Research Cluster ELEMENTS (Project ID 500/10.006).
D.H.R.~thanks the Galileo Galilei Institute for Theoretical Physics for the hospitality and the INFN for partial support during the completion of this work.
\end{acknowledgments}

\end{document}